\documentclass[12pt,preprint]{aastex}

\slugcomment{Submitted to Astrophysical Journal Letters}

\shorttitle{Echo from SN 2006X}
\shortauthors{Crotts et al.}

\begin{document}

\title{The Nature and Geometry of the Light Echo from SN 2006X\footnote{Based
on observations made with the NASA/ESA Hubble Space Telescope, obtained at the
Space Telescope Science Institute, which is operated by the Association of
Universities for Research in Astronomy, Inc., under NASA contract NAS5-26555.}}

\author{Arlin Crotts$^1$ and David Yourdon$^1$}
\affil{$^1$Dept.\ of Astronomy, Columbia University,
    550 W.\ 120th St., New York, NY~ 10027}
\email{email contact: arlin@astro.columbia.edu}

\begin{abstract}
We report on the discovery of the geometry producing the light echo emanating
from Supernova 2006X, a nearby but underluminous Type Ia in M100 (= NGC 4321).
This offers a rare chance to study the environment of a Type Ia supernova.
Contrary to previous reports, there is little evidence of a circumstellar
component in the light echo morphology or in the light curve of the unresolved
SN point source.
Instead, the obvious and dominant echo contribution comes from what is
probably a relatively thin sheet of material some 26 pc in front of SN 2006X.
Of the four known Type Ia light echoes, three show no evidence of a
circumstellar echo and the fourth needs to be confirmed.
We consider other evidence for circumstellar material around SN Ia, which may
be rare.

\end{abstract}


\section{Introduction}

Intense interest is focused on Type Ia supernovae (SNe Ia) as cosmological
standard candles.
To isolate cosmological parameters, several such complementary probes are
required, and SNe Ia compose the best proven, most powerful technique
(Kolb 2005), and in many ways are the easiest to realize at a level needed to
reveal dark energy.
It is often assumed that systematic errors in SN Ia photometry will improve
sufficiently for further refinement of cosmological tests to succeed; they must
acheive a high level of photometric precision and insensitivity to systematic
error in order to measure higher-order parameters characterizing dark energy
(see Linder \& Huterer 2003, Tonry 2004).

While several techniques incorporate extinction corrections to predict the
luminosity of SNe Ia, dust in the interstellar and circumstellar medium (ISM
and CSM) can have other effects.
For instance, Patat (2005) shows that light echoes from dust can introduce
several percent changes into the luminosity $L$, particularly on the decrease
from maximum light, and reducing $L$ while increasing maximum lightcurve width
$\Delta m_{15}$ in a stochastic manner depending on seemingly random placement
of surrounding material.

There are many ways in which SN Ia luminosity might relate to environment.
Hamuy et al.\ (2000) explore how SN Ia brightness is related to galaxian
integrated colors and speculate that this connection proceeds via metallicity.
Umeda et al.\ (1999) suggest that metallicity impacts the carbon fraction in
C$+$O white dwarf stars (WD).
Recent evidence points to a relation of $\Delta m_{15}$ to galaxy morphology
(Altavilla et al.\ 2004, Della Valle et al.\ 2005) and there appears to be
bimodality in progenitor age ($\approx 1$ Gy vs.\ a few Gy: Mannucci et
al.\ 2006), which is reflected in $\Delta m_{15}$ (Hamuy et al.\ 1996, van den
Bergh et al.\ 2005).
These effects are sufficiently well known to SN observers but become much more
difficult to treat at higher redshift where host galaxies are more ellusive
e.g., Foley et al.\ (2007).

There are several ways to determine if substantial circumstellar effects
influence SNe Ia e.g., high-velocity and time-variable spectroscopic components
(Gerardy et al.\ 2004, Mazzali et al.\ 2005, Wang et al.\ 2003, Quimby et
al.\ 2006) that might be explained alternatively (Quimby et al.\ 2006) as
structure internal to the explosion (Mazzali et al.\ 2005).
In many or most cases such efforts fail to establish a circumstellar component
e.g., SN 2000cx (Patat et al.\ 2007a), and when detected sometimes involves
peculiar SNe Ia (Panagia et al.\ 2006) or questionable (Benetti et al.\ 2006)
SNe Ia (see also Hamuy et al.\ 2003, Aldering et al.\ 2006, Prieto et
al.\ 2007).
Hamuy et al.\ (2003) suggest a new class of CSM-dominated Type Ia (for example
SNe 1997cy, 1999E and 2002ic), but these would compose fewer than 1\% of all
SNe Ia.
In general only upper limits are placed on the mass-loss rate by optical
(Mattila et al.\ 2005), radio (Panagia et al.\ 2006) and UV/X observations
(Immler et al.\ 2006).
Likewise, theory is a poor guide, in part because SN Ia progenitors are a small
and not well-isolated subcomponent of accreting WDs: SN Ia progenitors compose
probably about 15\% of binaries in the appropriate initial mass range
(3--8 $M_\odot$) and almost certainly 2--40\% of these (Maoz 2007).
Since the last known Galactic SN Ia was in the year 1604 (and last in the Local
Group probably in 1885), detailed ground-based study of SN Ia remnants or the
small fraction of Ia progenitors among the larger population of WD accreting
binaries is not definitive in terms of their CSM.

Light echoes might easily provide decisive clues regarding interstellar versus
circumstellar intervening material.
We can learn about the 3-D distribution of scattering material by taking an
image at time $t$ after maximum light, measuring the angular radius $r$ of an
echoing cloud (e.g., in light-years), and one can directly infer the cloud's
foreground distance.
Only three other SNe Ia have shown the clear evidence of light echoes (SNe
1991T, 1995E and 1998bu).
Here we report on the observations that allowed discovery of a fourth, in front
of SN 2006X, as well as details of those echoes regarding interstellar or
circumstellar origin.

\section{Observations}

SN 2006X is a rare, fortuitous case of a SN in a field imaged extensively by
$HST$ before the explosion.\footnote{with WFPC2 on 1994 January 7, with 1800s
using F439W, 1668s F555W, 2318s F702, and on 1996 July 29, with 2400s F439W,
800s F555W, 1000s F675W, 2600s F336W, 1200s F814W, and 3900s F658N - both from
GO-5195 by Sparks et al.}
Furthermore the extended tail in the B-band lightcurve of SN 2006X strongly
hinted at the presence of a light echo (Crotts \& Sugerman 2006).
We observed SN 2006X on three visits with $HST$ after explosion.
On UT 2006 May 21 we took a series of rapid exposures to avoid saturating the
SN (then at $V = 17.2$) to maintain PSF-fitting and image subtraction efficacy
(in total: 1480s F435W, 1080s F555W, 1080s F775W, in the ACS/HRC bands closest
to those from WFPC2 above).
We also took care to include in each image a bright, unsaturated stellar source
for PSF comparison.
The same bands were observed on UT 2006 December 24 (920s F435W, 520s F555W,
520s F775W).
By the time of our third visit, UT 2008 January 4, ACS was unavailable, so we
used the closest available bands with WFPC2 PC (1000s F380W, 1000s F439W, 2000s
F555W, 1000s F702W, 1000s F791W).
Throughout this paper we refer to epochs relative to $V$-band maximum on UT
2006 Feb 22.8 (Wang et al.\ 2008a), which for these $HST$ visits occur 87, 304
and 680 days post-maximum.

For our day 304 post-maximum epoch, the extension of the image of SN 2006X
beyond the PSF served as a strong clue of a light echo, since no such
nebulosity was evident at this position in the pre-SN images (Crotts 2007).
Wang et al.\ (2008b, hereafter W+08) show on the basis of these same $HST$
images the presence of extended nebulosity consistent with a light echo, and
confirm this with Keck/LRIS and DEIMOS spectra of this nebulosity similar to SN
2006X at maximum light, as might be expected by a light echo.
While a portion of this echo is resolved by $HST$, W+08 conclude on the basis
of the strength of the echo-like component in their spectra that there is
likely an additional component at smaller radii, probably circumstellar to the
SN progenitor.

\section{Morphological Analysis}

Our analysis of the 2006 May 21 (87d post-max) images does not indicate any
deviation from a stellar PSF at a statistically significant level.
The interstellar echo evident in later epochs (having $V = 21.96$) is too dim
to stand out evidently against the SN point source ($V = 17.23$).

In contrast, the 2006 December 24 (304d post-max) images show a significant
indication of a non-stellar source, which motivated our 2008 January 4
observations (860d post-max).
When we subtract a maximal-brightness point-source from these images, this
produces a circular residual with a significant depression of flux markedly
similar to that found by W+08 in their analysis of our images, so we will not
present this again here.

The quality of information for the interstellar echo in the 860d post-max data
is, however, sufficiently superior that many features questionable in the 2006
data are dealt with clearly.
A roughly ring-like source is seen in all five WFPC2 bands, consistent with
the same morphology throughout.
We present the scaled median average from the redder three of the five bands
(F555W, F702W and F791W, with 4, 2 and 2 sub-exposures apiece, summed by a
CR-split algorithm) in Fig.\ 1.
The pixel brightnesses actually increase moving out from the SN position out to
a radius of 0.075 arcsec.
The ring is not uniformly bright; the peak in brightness occurs at PA
50$^\circ$.

By convolving an infinitesimally-thin annular ring of constant radius (but not
constant azimuthal brightness profile) with the PSF and minimizing the
resultant $\chi ^2$ of flux in the affected pixels once this ring is subtracted
from the original image, we find a ring radius of 1.65 $\pm$ 0.1 pixels $=$
0.075 $\pm$ 0.005 arcsec.
At a distance to M100 of 15.2 Mpc (Freedman et al.\ 2001), this corresponds to
a radius of 5.5 pc (or 18.0 light-years).
Since the foreground distance $z$ in front of the SN of light echoing material
a radius $r$ (perpendicular to the sightline) at time $t$ after SN maximum
light is given by 

$$z = r^2/2ct - ct/2$$

\noindent
which for this echo implies $z = 85.7 \pm 10.3$ light-year (or 26.3 pc) at
$t = 1.87$~y after SN maximum light, with an average scattering angle of
12$^\circ$.
The natural FWHM of this interstellar echo annulus, due to the duration of the
maximum light peak, is at least 0.6 light-year, and probably thicker due to the
finite thickness of the reflecting dust layer, but we cannot constrain this at
a useful level due to the angular resolution limits of $HST$/WFPC2 and the
limited $S/N$ of the echo image.
The spread in radius is fit poorly with a FWHM more than about 20\% of the
central value, hence a spread in $z$ of more than 40\%, corresponding to
16 pc $\la$ $z$ $\la$ 36 pc.

There is little evidence in the 680d post-max morphology for a unresolved
circumstellar source strongly suggested by W+08.
The pixel containing the SN position contains 3\% of the total flux from the
entire echo nebula (in the redder three bands), but 21\% of any hypothetical
point source there.
However, much brighter pixels centered 1-2 pixel widths away imply that 69\% of
this flux must come from a source outside the central pixel.
The upper limit (3$\sigma$) on a circumstellar source is therefore about 5\% of
the total flux of the nebula.

\section{Lightcurve Analysis and Multiband Photometry}

We consider the lightcurve in the band in which the echo is best detected,
F555W, which is common to both ACS and WFPC2 images.
For day 87 post-max we detect no deviation (over $1\sigma$) from the the
stellar PSF for SN 2006X, so adopt the entire flux for the SN point-source
value, $V=17.23$.
For day 680 post-max, the echo nebulosity has $V=21.96$ of which a central
point source is less than 5\%, or $V > 25.2$, implying that the extended echo
itself has $21.96 < V < 22.02$.
In the case of day 304 post-max, the SN$+$echo signal corresponds to $V=20.53$.
Subtracting the SN central point source yields $V \approx 20$, to within
$\sim$0.1 mag error due to subtraction systematics, and hence a SN $V=20.88$.
This echo magnitude is in good agreement with the 680d post-max echo, as well
as that found by W+08 for 304d post-max.
There is no evidence of any change in the echo brightness between 304d and 680d
post-maximum; indeed the data is inconsistent with any change larger than
$\sim$10\%.
These photometric results are summarized in Fig.\ 2, along with various
comparison $V$ values from ground-based photometry of SN 2006X and other normal
SNe Ia.
Assuming that normal SNe Ia have no significant circumstellar echoes 304 or
670d after maximum, there is also no evidence that SN 2006X does, either.

We have compiled a library of many epochs of SN 2006X spectra taken at MDM
Observatory (Crotts et al.\ 2008) and use this to compare synthetic magnitudes
for the fluence of SN light in the $HST$ bands to the actual echo flux seen in
these bands.
A best fit reflectance function to the extended echo, for the F380W, F439W,
F555W, F702W and F791W bands, corresponds to $\lambda ^{-1.7}$ for a power-law
in wavelength (with an error of $\pm 0.2$ in the exponent).
Such reflectance wavelength dependence is consistent with Galactic dust of
$0.0005 - 2\mu$m radius (e.g., Sugerman 2003) but not smaller dust (large cut
off at 0.01$\mu$m or 0.1$\mu$m).
The echo has exceptionally high surface brightness ($\mu_V < 18.5$), indicating
$A_V$ values through the echoing cloud of at least a few (as per Sugerman 2003,
assuming MRN-like dust, and depending on the exact scattering phase function
adopted).
Wang et al.\ (2008a) estimate the host extinction value as $A_V = 2.24$ mag,
so this strongly suggests that this interstellar echo arises from the primary
source of extinction along the line of sight.



%

Magellan/Mike 6-km s$^{-1}$ resolution spectra of SN 2006X taken on Feb.\ 13.35
and 23.25 UT show unusually strong absorption lines in Na I, Ca II, K I and the
CN B-X (0,0) violet bands (Lauroesch et al.\ 2006, see also Patat et al.\ 2007,
hereafter P+07, for comparable data).
The Na I $\lambda$5889 equivalent width is $W_{eq} = 0.74$\AA\ and heavily
saturated (and composed of at least two components, due to line asymmetry), and
the CN R(0) 387.46-nm line has an equivalent width (0.09\AA), stronger than any
published strength through Galactic interstellar clouds.
Based on Galactic expectation e.g., Crutcher (1985), one should expect
$A_V > 3$.
These lines (at $\sim 1630$ km s$^{-1}$) are redshifted 72 km s$^{-1}$ versus
the M100 centroid, consistent with the value of the galaxy's rotation curve at
this point.
The echo, absorption strengths and velocities of these features are all
consistent with a single origin in the galaxy's disk, suggesting that the
SN 2006X progenitor sits $\sim$26 pc behind the disk.

\section{A Circumstellar Component?}

There are several pieces of evidence suggesting a circumstellar absorption-line
component or echo feature in front of SN 2006X, most notably time-variable Na I
absorption lines in high-resolution spectroscopy (P+07), which we discuss
momentarily.

Our analysis above places strong limits on the brightness of any circumstellar
echo 680d post-max; furthermore is shows that there is no evidence of a
circumstellar echo 304d post-max, and implies that any such echo likely has
$V > 24.5$.
On the basis of the difference between the resolved echo brightness 304d
post-maximum and a SN maximum-like spectrum detected one month earlier (274d
post-max), W+08 claim a circumstellar component to the echo which
is $0.6 \pm 0.3$ times as bright as the $V = 22$ interstellar echo.
At 680d after maximum, any such component must have dimmed to $V > 25.2$, or at
least 2.6 mag fainter.
Even if the echo paraboloid encounters no material after 274d post-max,
the echo could dim no more than 8 mag by 680d post-max.
(One can assume that the echoing dust, at least $\sim 10^{18}$ cm from the SN,
is not destroyed.)~
Thus the circumstellar material might extend at undiminished density beyond the
274d$=$ 0.75y paraboloid by another 0.6y, corresponding to a total radius of
0.68 light-year for material directly behind the SN as seen from Earth, or
1.35 light-year (=$1.3\times10^{18}$ cm) radius, 90$^\circ$ to one side.
The detection of a circumstellar component (W+08) is only a $2\sigma$ result,
however, and this confidence estimate does not include several likely
significant sources of systematic error.
There is no evidence from the $HST$ imaging sample that supports the
existence of a circumstellar echo, and indeed this component is limited to a
value of only $\sim$10\% of the total echo at day 304 post-maximum, not the
60\% ($\pm$30\%) reported by W+08 for day 274.

P+07 describe $\sim 6$ km s$^{-1}$ resolution spectroscopy of SN 2006X on days
$-$5, $+$11, $+$58, $+$102 and $+$118 with respect to V-band maximum.
At velocities up to $\sim$150 km s$^{-1}$ blueshifted from the strong
absorption line described above arise several weak Na I components that are
highly variable in structure over the first three epochs.
P+07 explain this variability as Na I ionized by the initial SN flash.
The fact that Ca II remains unchanged during all of these epochs is explained
by its higher ionization potential.
Somewhat paradoxically, while most of these Na I components increase with time,
those with $15 < v < 55$ km s$^{-1}$ decrease after $+$11d post-maximum.
P+07 theorize that these components are destroyed by SN ejecta, so are at
$\sim 10^{16}$ cm from the SN.
Ionization strength arguments place all time-dependent lines within $few \times
10^{16}$ cm.
In contrast, Chugai (2008) calculates that the time-variable Na I absorption
components must arise at least $10^{17}$ cm from SN 2006X, since there is no
evidence of grain destruction as would be traced in Ca II absorption-line
variation.

No radio emission has ever been detected from SN 2006X, for several
observations spanning from two days to two years after the discovery date
(UT 2006 February 7.1) and frequencies 4.88 GHz - 22.46 GHz (Stockdale et
al.\ 2006, Chandra et al.\ 2006, 2008).
Scaling from Panagia et al.\ (2006), this limits correspond to mass-loss rates
of the smaller than $10^{-8}$ to a $few \times 10^{-8}$ M$_\odot$ y$^{-1}$,
comparable to the limits Chugai (2008) places based on the absence of Ca II
absorption components not also seen in Na I at $10^{-8}$ M$_\odot$ y$^{-1}$ for
an outflow velocity of 10 km s$^{-1}$.

Both P+07 and Chugai (2008) discuss the relative merits of a circumstellar
versus interstellar origins of the time-variable Na I absorption components,
especially considering the possible small-scale structure that might arise in
such a thick interstellar cloud as the one seen at 26 pc distance, due to
motion of the source against foreground absorption.
The imaging and photometric data presented here has little constraint on such
ISM structure (at the $\sim 10^{13} - 10^{15}$ cm level), and if these
absorption lines are circumstellar, they arise at radii $\sim 10^{16} -
10^{18}$ cm where the light echo data similarly have little bearing.

\section{Conclusions}

In conclusion, the $HST$ imaging 87, 304 and 680d after V-band maximum say
little about circumstellar material closer than about $1.3\times 10^{18}$ cm.
As in the case of SN 1991T and 1995E, no central source corresponding to a
circumstellar structure, echoing or not, has ever been detected.
In contrast, SN 1998bu shows a central source that has been interpreted as a
circumstellar echo.
We note, however, that an image-subtraction comparison of the existing archival
$HST$ imaging shows very little change in this source between years 2000 and
2006.
We are investigating if spectroscopy of this source will reveal it as an echo,
a supernova remnant, or some other structure (Crotts 2008).

To summarize, there are a number of significant results that these new
observations have produced:

1) The echo signal arises primarily from a presumably interstellar sheet of
material 26.3$\pm$3.2 pc in the SN foreground, and this structure likely
represents the dominant portion of the large amount of extinction along the SN
sightline.

2) As of 680 days after V-band maximum light there is no indication of any
circumstellar echo (at any level above 5\% of the total echo signal), and the
imaging and photometry at 304 days is consistent with these results, both in
the presence of an interstellar echo of similar geometry and brightness, and
the absence of evidence of a circumstellar echo.

3) There is no evidence of any circumstellar echo signal at any epoch in these
data, but they do not bear directly on the plausibly circumstellar absorption
signal seen in Na I making SN 2006X a candidate for one of the few Type Ia
supernovae in which circumstellar matter has been knowingly detected.

\acknowledgments
I thank Steve Lawrence and Ben Sugerman for helpful discussion.
This work was supported by grants GO/DD-10991 and GO 11171 from STScI.

\noindent
{\bf References:}

\noindent
Aldering, G., et al.\ 2006, ApJ, 650, 510.

\noindent
Altavilla, G., et al.\ 2004, MNRAS, 349, 1344.

\noindent
Benetti, S., et al.\ 2006, ApJ, 653, L129.

\noindent
Chandra, P., Chavalier, R.\ \& Patat, F.\ 2006, ATel, 454.

\noindent
Chandra, P., Chavalier, R.\ \& Patat, F.\ 2008, ATel, 1393.

\noindent
Chugai, N.N.\ 2008, Astron.Let., in press (arXiv:0801.4468).

\noindent
Crotts, A.P.S.\ 2008, in preparation.

\noindent
Crotts, A.P.S.\ 2007, $HST$ proposal 11171
(http://www.stsci.edu/observing/phase2-public/ 11171.pro).

\noindent
Crotts, A.P.S.\ \& Sugerman, B.E.\ 2006, $HST$ proposal 10991
(http://www.stsci.edu/observing/ phase2-public/10991.pro).

\noindent
Crotts, A.P.S., et al.\ 2008, in preparation.

\noindent
Crutcher, R.M.\ 1985, ApJ, 288, 604.

\noindent
Della Valle, M., Panagia, N., Padovani, P., Cappellaro, E., Mannucci, F.\ \&
Turatto, M.\ 2005, ApJ, 629, 750.

\noindent
Foley, R.J., et al.\ 2007, ApJ, submitted (preprint at arXiv:0710.2338).

\noindent
Freedman, W.L., et al.\ 2001, ApJ, 553, 47.

\noindent
Gerardy, C., et al.\ 2004, ApJ, 607, 391.

\noindent
Hamuy, M., et al.\ 1996, AJ, 112, 2348.

\noindent
Hamuy, M., et al.\ 2000, AJ, 120, 1479 (Erratum, 122, 3506).

\noindent
Hamuy, M., et al.\ 2003, Nature, 424, 651.

\noindent
Immler, S.I., et al.\ 2006. ApJ, 648, L119.

\noindent
Kolb, E.\ 2005, report of Dark Energy Task Force to the NSF-NASA-DOE Astronomy
and Astrophysics Advisory Committee.

\noindent
Lauroesch, J.T., Crotts, A.P.S., Meiring, J., Kulkarni, V.P., Welty, D.E.\ \&
York, D.G.\ 2006, CBET, 421.

\noindent
Lair, J., Leising, M.D., Milne, P.A.\ \& Williams, G.G.\ 2006, AJ, 132, 2024.

\noindent
Linder, E.V.\ \& Huterer, D.\ 2003, Phys.\ Rev.\ D, 67, 081303.

\noindent
Mannucci, F., Della Valle, M.\ \& Panagia, N.\ 2006, MNRAS, 370, 773.

\noindent
Maoz, D.\ 2008, MNRAS, 384, 267.

\noindent
Mattila, S., et al.\ 2005, A\&A, 443, 649.

\noindent
Mazzali, P.A., et al.\ 2005, MNRAS, 357, 200.

\noindent
Panagia, N., et al.\ 2006, ApJ, 646, 349.

\noindent
Patat, F.\ 2005, MNRAS, 357, 1161.

\noindent
Patat, F., et al.\ 2007a, A\&A, 474, 931.

\noindent
Patat, F., et al.\ 2007b, Science, 317, 924. (P+07)

\noindent
Prieto, J.L., et al.\ 2007, ApJ, submitted (preprint at arXiv:0706:4088).

\noindent
Quimby, R., et al.\ 2006, ApJ, 636, 400.

\noindent
Stockdale, C.J., et al.\ 2006, CBET, 396.

\noindent
Sugerman, B.E.K.\ 2003, ApJ, 126, 1939.

\noindent
Tonry, J.L.\ 2005, Phys.\ Scr., T117, 11.

\noindent
Umeda, H., Nomoto, K., Yamaoka, H.\ \& Wanajo, S.\ 1999, ApJ, 513, 861.

\noindent
van den Bergh, S., Li, W.\ \& Filippenko, A.\ 2005, PASP, 117, 773.

\noindent
Wang, L., et al.\ 2003, ApJ, 591, 1110.

\noindent
Wang, X., et al.\ 2008a, ApJ, 675, 626.

\noindent
Wang, X., et al.\ 2008b, ApJ, in press (arXiv:0711.2570). (W+08)


\clearpage

\begin{figure}
\plotfiddle{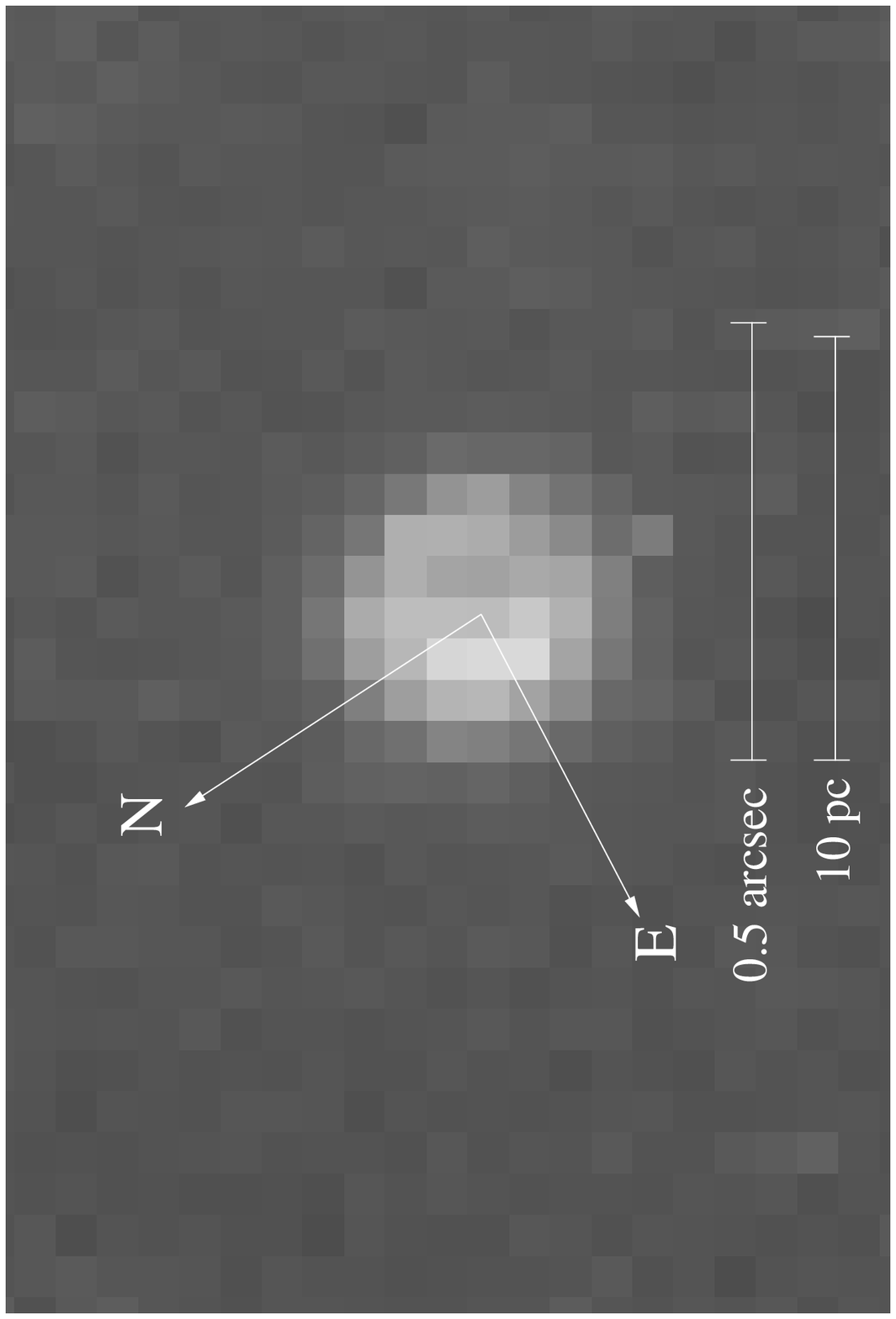}{0.0in}{-90}{400}{600}{-070}{-1417}
\vskip -2.0in
\caption {
A very recent WFPC2 PC image (UT 2008 January 4, 680d post-maximum) of SN
2006X, showing an interstellar echo in excellent agreement with a cloud 26 pc
in front of the SN in the ISM.
This represents 4000s of exposure, summing the median-summed images taken in
F555W, F702W, and F791W bands.
The echo has the shape of a narrow ring 11 pc in diameter, which appears
to be expanding at about $4c$ since 304d post-maximum, consistent for the echo
expansion velocity expected for a plane 26 pc in the SN foreground (5.8$c$).
The presence of a CSM source in the center is ruled out any level above 5\% of
the echo total flux, to $3\sigma$ confidence.
The echo's brighness is consistent with the observed SN extension seen on 2006
December 24.
The box is 1.48 arcsec across, and we maintain the original image rotation
(with vertical at PA$= - 34^\circ$, with North up and slightly left, and East
left and slightly down).
Scale is shown by angular and physical reference bars.
}
\end{figure}

\begin{figure}
\plotfiddle{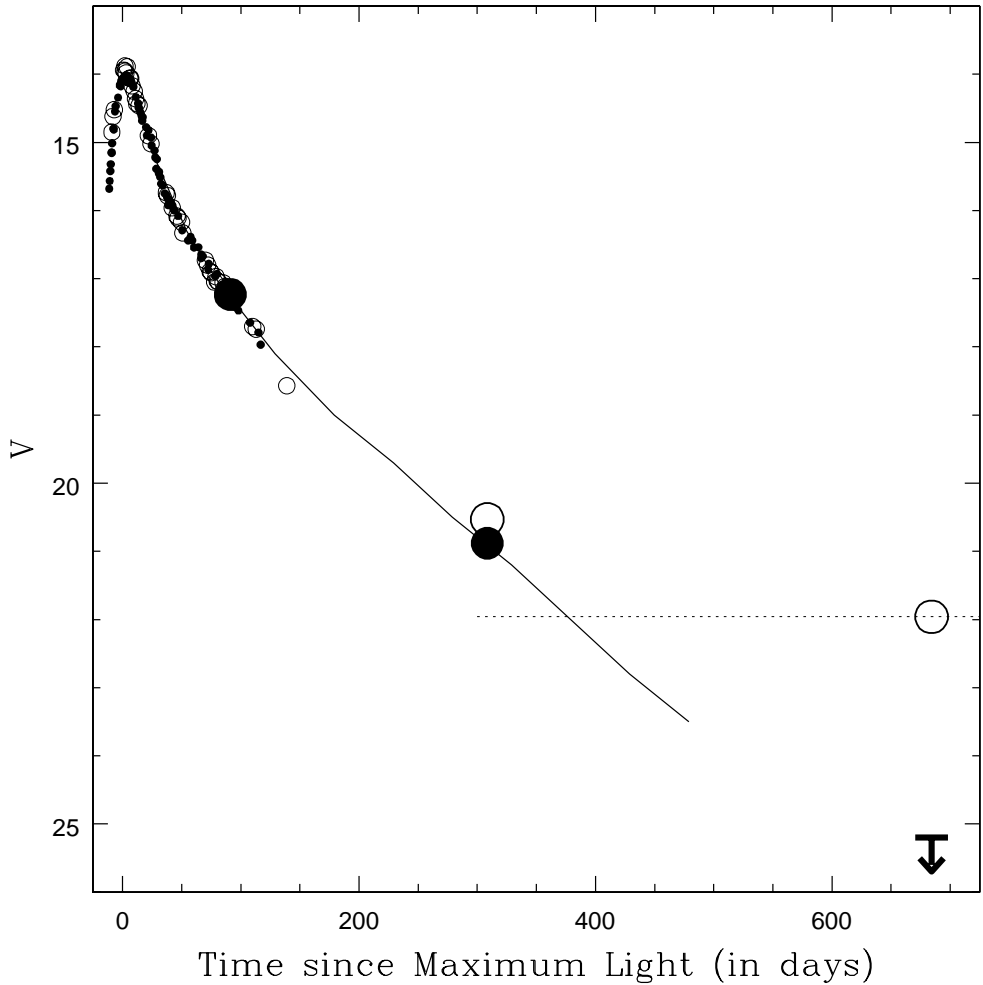}{-0.8in}{000}{400}{400}{+020}{-217}
\caption {
V-band light curve of SN 2006X from various sources.
The numerous smaller, filled circles indicate ground-based photometry from Wang
et al.\ 2008a, and the smaller of the two sets of open circles represent
ground-based photometry from MDM Observatory (Crotts et al.\ 2008).
The solid curve indicates the average of seven normal SN Ia light curves (SNe
2000E, 2000ce, 2000cx, 2001C, 2001V, 2001bg and 2001dp, over days 100 to 500,
from Lair et al.\ 2006, adjusted for distance and extinction.
Superimposed are the detected point-sources or $3\sigma$ upper limit for
SN 2006X plus any circumstellar component taken from $HST$ imaging (large solid
circles and upper limit).
Note that the upper limit value for the central point source 680d post-maximum
is about a magnitude or more above a linear extrapolation of the Lair et
al.\ curve.
These values for the SN point source are distinguished from the total source
brightness for SN 2006X (large open circles), which presumably are different
due to light echo contributions.
The echo magnitude seen 680 days after maximum is indicated by the horizontal
dashed line, consistent with the echo seen 304 days post-maximum.
}
\end{figure}

\end{document}